\documentclass[showpacs,bibnotes]{revtex4}
\usepackage{graphicx}
\begin{document}
\draft
\def\ds{\displaystyle}
\title{Feshbach Resonance and Growth of a Bose-Einstein Condensate}
\author{C. Yuce }
\email{cyuce@anadolu.edu.tr}
\author{A. Kilic}
\address{ Department of Physics, Anadolu University,
 Eskisehir, Turkey}
\date{\today}
\begin{abstract}
Gross-Pitaevskii equation with gain is used to model Bose Einstein
condensation (BEC) fed by the surrounding thermal cloud. It is
shown that the number of atoms continuously injected into BEC from
the reservoir can be controlled by applying the external magnetic
field via Feshbach resonance.
\end{abstract}
\maketitle
\section{Introduction}

Gross-Pitaevskii (GP) equation is one of the most important
nonlinear equations used in physics. It appears in many branches
of physics such as nonlinear optics, fluid mechanics, condensed
matter and theory of superfluidity. For example, GP equation is
used to explore the macroscopic behavior of the Bose-Einstein
condensation (BEC) at low temperature. Its validity in the BEC is
based on the condition that the s-wave scattering length should be
much smaller than the average distance between the atoms and that
the number of atoms should be much larger than one. Experimental
achievements and theoretical studies of Bose-Einstein condensates
(BEC) of weakly interacting atoms have stimulated intensive
interest in the field of atomic matter waves and nonlinear
excitations such as dark \cite{dark1,dark2,dark3} and bright
solitons \cite{bright1,bright2,bright3,bright4}. The dynamics of
BEC crucially depends on the sign of interatomic interaction. Dark
(bright) solitons that are solutions to the GP equation can be
generated with repulsive (attractive) interactions, resulting from
the positive (negative) s-wave scattering length. In the present
study, we start by considering the Gross-Pitaevskii (GP) equation
modified by a linear term. The Gross-Pitaevskii gain equation
reads;
\begin{equation}\label{denklemoncesi}
i\hbar \frac{\partial \Psi}{\partial t}=-\frac{\hbar^2}{2m}
\nabla^2 \Psi+V_{trap}\Psi+\frac{4\pi \hbar^2}{m}a_s |\Psi|^2
\Psi+i \hbar \frac{\Gamma }{2} \Psi~,
\end{equation}
where
$\ds{V_{trap}=\frac{m}{2}\omega_{\perp}^2(x^2+y^2)+\frac{m}{2}\omega_{z}^2
z^2}$, $a_s(t)$ is the time-dependent scattering length and the
constant $\Gamma$ is the gain (loss) term for $\Gamma
>0$ $(\Gamma <0)$. The complex linear term with positive $\Gamma$ accounts for the mechanism
of loading the thermal cloud into the BEC by optically pumping
atoms from the external cold atomic-beam source. From the physical
point of view, replenishing is required in steady state to
compensate for various intrinsic loss mechanisms, such as
collisions with hot atoms from the background vapor and inelastic
two and three-body collisions between the trapped atoms. However,
for the negative values of $\Gamma$, the above equation describes
a BEC that is continuously depleted by loss.\\
For attractive interaction, the number of atoms in the condensate
grows until the total number of atoms in the condensate exceeds a
critical value with which the BEC undergoes collapse. During the
collapse, the number of atoms is decreased. If the condensate is
fed by a surrounding thermal cloud, then the condensate undergoes
cycles of growth and collapses. Hulet and his team at Rice
university observed the growth of a condensate of trapped $^7 $Li
atoms with attractive interaction and its subsequent collapse
\cite{gerton}. Ketterle with his team measured the rate of growth
of a $^{23}$Na condensate fed by a thermal cloud \cite{miesner}.
Growth of a Bose-Einstein condensate from thermal vapor was also
experimentally realized for atoms of $^{87}$Rb \cite{2kohl}. Some
methods have been introduced theoretically to account for the
growth of the BEC. GP gain equation was used to model the growth
of a BEC by Drummond and Kheruntsyan \cite{drummond}. They
determined that, as the condensate grows, the center of mass
oscillates in the trap.\\
Furthermore, GP gain equation is of special importance in the
field of an atom laser, that is a device which produces an intense
coherent beam of atoms by a stimulated process
\cite{generic, generic2}.\\
In recent years, GP gain equation (\ref{denklemoncesi}) has also
been studied mathematically by some authors
\cite{atre,serkin,kruglov}. GP gain equation can also be used in
the theory of optical fibers if the harmonic trap potential is
assumed to be
zero \cite{serkin,kruglov}.\\
The interaction effect in the system (\ref{denklemoncesi}) is
determined by the s-wave scattering length. In ref.
\cite{feshbach1}, it has been shown that controlling the
generation of bright and dark soliton trains from periodic waves
can be achieved by the variation of the scattering length. Recent
experiments have demonstrated that tuning of the s-wave scattering
length can be achieved due to the Feshbach resonance
\cite{fesh1,fesh2,fesh3,fesh4,fesh5}. It offers a possibility to
vary the interaction strength in ultracold atomic gases simply by
applying an external magnetic field.
\begin{equation}\label{formula1}
a_s(t)=a \{ 1+\Delta / [B_0-B(t)]    \} ~,
\end{equation}
where ${\ds a}$ is the asymptotic value of the scattering length
far from resonance, ${\ds B(t)}$ is the time-dependent externally
applied magnetic field, $\Delta$ is the width of resonance and
${\ds B_0}$ is the resonant value of the magnetic field. The
strength of the interaction can be adjusted experimentally from
large negative values to large positive values. As a result, the
experiments with magnetically trapped ultra cold atomic gases,
where the s-wave scattering length fully determines the
interaction effects, have an unprecedented high level of control
over the interatomic interactions. With this experimental degree
of freedom, it is
possible to intensively study the dynamics of BEC.\\
Here, we will study the changes in the number of atoms supplied by
a surrounding thermal cloud due to the variations of the
scattering length by the external magnetic field. We will assume
that $\Gamma$ does not change but
the s-wave scattering length changes via Feshbach resonance.\\
By supposing the number of particles is not very large for the
cigar-shaped traps, it is legitimate to use the reduced 1D GP
equation
\begin{equation}\label{denklem}
i\frac{\partial \Psi}{\partial t}=-\frac{1}{2} \frac{\partial^2
\Psi}{\partial z^2}+\frac{\omega^2}{2} z^2 \Psi+\frac{2a_s}{a_B}
|\Psi|^2 \Psi+i \frac{\gamma }{2} \Psi~,
\end{equation}
where $\ds{\omega^2= \omega_z^2/\omega_{\perp}^2 }$,
$\ds{\Gamma=\gamma/\omega_{\perp}}$ and $\ds{a_B}$ is the Bohr
radius. Here, the coordinate $z$ and the time $t$ are measured in
units of $a_\perp$ and $1/\omega_\perp$, respectively. The
relationship between the macroscopic wave functions
$\ds{\Psi(\vec{r},t)}$ and $\ds{\Psi(z,t)}$ is given by
\begin{equation}
\Psi(\vec{r},t)=\frac{1}{\sqrt{2 \pi a_B} a_{\perp}} \exp{(-i
\omega_{\perp} t-\frac{x^2+y^2}{2 a_{\perp}^2})} \times
\Psi(\frac{z}{a_{\perp}}, \omega_{\perp} t) ~,
\end{equation}
where $\ds{a_{\perp}=(\hbar/m \omega_{\perp} )^{1/2}}$. To extract
qualitative information and to improve the understanding of the
underlying physics, we will transform the GP gain equation
(\ref{denklem}) analytically to the one with
the constant scattering length.\\
To get rid of the time-dependent scattering length from the
equation (\ref{denklem}), the two transformations are introduced
as follows;
\begin{equation}\label{idfd}
\Psi(z,t)\rightarrow {f(t)}^{-1/2}~\exp{\left(\int \frac{
\gamma(f^2-1)}{f^2} dt^{\prime} +\frac{i\dot{f} }{2 f}~z^2\right)}
\Phi(z,t);~~~~~~z\rightarrow f(t) ~z~,
\end{equation}
where dot denotes time derivation and the time-dependent function
$f(t)$ is given by
\begin{equation}\label{slozel}
f(t)= \sqrt{\frac{ \sin 2\omega t +\sqrt{1+\Omega^2}}{\omega}}~,
\end{equation}
where $\Omega$ is a constant. Note that the time-dependent
function $f(t)$ is chosen in such a way that it satisfies $\ds{
f^3 \ddot{f}+\omega^2 f^4=\Omega^2 }$. The scale transformation on
the coordinate accounts for the contraction or the expansion of
the condensate depending on the behavior of $f(t)$. The
time-dependent scaling of the coordinate and re-scaling of the
wave function have been previously applied to harmonic oscillator
potential with the time-dependent angular frequency $\omega^2(t)$
\cite{ekl1, ekl2, ekl3}. These scale transformations are used here
to explore
the effect of Feshbach resonance on the number of atoms injected into BEC.\\
Now, let us first substitute the transformation for $\Psi$ and
then for the coordinate into the equation (\ref{denklem}). Note
that under the coordinate scaling, the time derivative operator
transforms as $\ds{\partial/\partial t \rightarrow
\partial/\partial t- (\dot{f}/f) z~ \partial/\partial z}$. We
observe that the equation (\ref{denklem}) can be reduced to the
another GP gain equation with the constant scattering length
\begin{equation}\label{denklem2}
i \frac{\partial \Phi}{\partial \tau}=-\frac{1}{2}
\frac{\partial^2 \Phi}{\partial z^2}+\frac{\Omega^2}{2}
z^2\Phi+\frac{2a}{a_B} ~|\Phi|^2 \Phi+i \frac{\gamma }{2}\Phi~,
\end{equation}
where $a$ is a constant scattering length. We made another
transformation on time $\ds{\tau=\int_0^t f^{-2} dt^{\prime}}$ in
the last step. The necessary and sufficient condition for the
existence of such a reduction is given by
$\ds{a_s(t)=a~f^{-1}~\exp{\left(-2\int \frac{ \gamma(f^2-1)}{f^2}
dt^{\prime}\right)} }$. As a special case, if $\ds{ f^2 >> 1}$ for
some special values of $\ds{\omega^2}$ and $\ds{\Omega^2}$, then
$a_s(t)$ can be
approximated as $\ds{a_s(t)\approx a~f^{-1}e^{-2\gamma t} }$.\\
The magnitudes of trapping frequencies $\ds{(\Omega, \omega)}$ are
scaled. We have freedom to make them equal by choosing
$\ds{\Omega=\omega}$ or to turn-off the trap by
choosing $\ds{\Omega=0}$.\\
It is interesting to observe that the equation (\ref{denklem}) can
be transformed into the another GP gain equation with a different
time parameter (\ref{denklem2}). The two equations
(\ref{denklem},\ref{denklem2}) describe the two different
condensates continuously replenished by pumping from the
reservoir. The only physical difference between the two
condensates is that the latter one is with the constant scattering
length while the former one is with Feshbach resonance. The
transition from one to the other is given by the
transformations (\ref{idfd}). \\
This reduction is of great importance in the sense that although
the corresponding exact solutions are not known, it allows us to
understand the effect of Feshbach resonance on the dynamics of the
condensates.\\
Let us find how the Feshbach resonance changes the number of atoms
supplied to the system from the reservoir by using (\ref{idfd}).
Note that the gain coefficient $\gamma$ doesn't change for the two
condensates.
\begin{equation}\label{denklem51}
\int |\Psi|^2 d^3 r=\exp{\left(\int \frac{ \gamma(f^2-1)}{f^2}
dt^{\prime}\right)} \int |\Phi|^2 d^3 r~.
\end{equation}
This formula specifically shows the effect of the time-dependent
nature of the scattering length on the number of atoms in the
condensates. The relation (\ref{denklem51}) has been derived
without obtaining the exact solution to the GP gain equation. It
is concluded that simply by applying the external magnetic field,
one can control the number of atoms transferred from the
surrounding cloud. In an experiment for the negative sign of
$\gamma$, one can also take advantage of Feshbach resonance to
control the number of atoms left in the BEC.\\
The method applied here is analytical and only works for a special
$a_s(t)$. Numerical calculations can be performed for the most
general $a_s(t)$ on the basis of the fact that Feshbach resonance
can be used to control
the number of atoms supplied to the system.\\
As a special case, if we set $\ds{\gamma=0}$, then the number of
atoms doesn't change for the two BECs
(\ref{denklem},\ref{denklem2}) as expected. In other words,
Feshbach resonance does not change the number of atoms if there is
no reservoir. The transformations (\ref{idfd}) conserve the total
number of atoms. However, the presence of $\ds{ f(t)}$ in the
transformations (\ref{idfd}) has the effect of increasing the
central peak of the condensate with a consequent
contraction of the atomic cloud. \\
If $\gamma \neq 0$, the number of particles as well as the
condensate size is changed by Feshbach resonance as can be seen
from the equation (\ref{denklem51}). The transformations (5) allow
us to understand the underlying physical effects of time-dependent
nature of the scattering length on GP gain equation, which is used
to model the condensate growth and atom laser.\\
Let us study the evolution of the system if we choose $\Omega=0$.
In this case, the equation (\ref{denklem2}) describes a BEC with
the trap potential turned-off and continuously replenished by
pumping from the reservoir. However, the equation (\ref{denklem})
describes another BEC whose scattering length changes with time
through the application of the external magnetic field. Atoms are
injected into both condensates by the same gain coefficient
$\gamma$. Since the harmonic potential does not exist and the
scattering length is constant in (\ref{denklem2}), its solution
can be constructed in terms of the well-known bright and dark
solitons depending on the sign of the scattering length. Then, the
exact solution for (\ref{denklem}) can also be constructed by
using the analytical
transformations given in (\ref{idfd}).\\
The time parameter transforms as $\ds{\tau=\frac{\sin \omega
t}{\sin \omega t+\cos \omega t}}$ when $\Omega=0$. The
transformations (\ref{idfd}) works in the time interval $0\leq
\omega t < 3\pi/4 $. In this interval, $\ds{f(t)=\sqrt{(1+\sin
2\omega t)/\omega}}$ varies from $\sqrt{1/\omega}$ to zero.
Switching-off the trap potential has dramatic effects on the
dynamics of the system. The longitudinal width of the two
condensates can be compared by making use of the coordinate
transformation $\ds{(z \rightarrow z~\sqrt{(1+\sin 2 \omega
t)/\omega})}$. If it is equal to $L$ for the BEC described by
(\ref{denklem2}), then it is equal to $\ds{L~ \sqrt{(1+\sin 2
\omega t)/\omega} }$ for the other BEC described by
(\ref{denklem}). The time $\ds{\omega t = \pi/4}$ is a turning
point in the sense that the longitudinal width of the BEC starts
to decrease from this time while it is increasing in the interval $0\leq\omega t\leq\pi/4$.\\
The condensate collapses into itself at time $\ds{\omega t=3
\pi/4}$ since the longitudinal width of the BEC (\ref{denklem})
goes to zero. Furthermore, the density of the BEC shows asymptotic
behavior depending on the sign of $\gamma$. At time $\ds{\omega t
=3 \pi/4}$, $\ds{|\Psi|^2 \rightarrow \infty |\Phi|^2}$ for
$\gamma >0$ while the longitudinal width of the BEC
(\ref{denklem}) approaches zero. The matter wave function becomes
somewhat like a Dirac delta function. This is reasonable, since
atoms are constantly injected into the condensate. However, this
is not physical, because the BEC becomes a singular point. From
the physical point of view, GP gain equation is not valid for the
description of the BEC if the density of the BEC is big enough. To
prevent the formation from being a true singularity, the
three-body recombination losses are expected to cause rapid
depletion of the BEC. In order to model atom loss due to 3-body
recombination, a phenomenological term proportional to $\Psi^4$
should be added to the GP gain equation (1) for a good physical
description of the BEC. Note also that, the scattering length $a_s
(t)$ goes to zero if $\gamma
>0$  at this time. Physically, the scattering length can be made zero by tuning the
external magnetic field. For example, at a magnetic field of ${\ds
B \approx 166.5 }$ G, the scattering length vanishes for
$\ds{{}^{85}}$Rb and the gas behaves effectively as an ideal Bose
gas. However, for the negative values $(\gamma <0)$, $\ds{|\Psi|^2
\rightarrow 0 |\Phi|^2}$ at time $\ds{\omega t =3 \pi/4}$ . Since
the atoms are lost for the negative values of $\gamma$, both the
longitudinal width and the number of atoms get smaller in time.
The scattering length $a_s (t)$ goes to infinity
if $\gamma <0$  at this time.\\
Let us now study the case if $\ds{\Omega^2 >0}$. The equations
(\ref{denklem},\ref{denklem2}) describe two condensates in the
confining harmonic traps with two different trap frequencies
$\omega^2$, $\Omega^2$. The scattering length is time-dependent
for the former, while it is constant for the latter. Both of them
are continuously replenished by pumping from the reservoir with
the same gain coefficient.\\
The relation between the time parameters is given by $\ds{\Omega
\tan \Omega \tau=1+\sqrt{1+\Omega^2}~\tan \omega t }$, where
$0\leq \omega t < \pi/2 $. $\ds{f(t)}$ is increasing in the
interval $\ds{0\leq t \leq \pi/2}$ and decreasing in the interval
$\pi/2 <t <\pi/4$. In this case, $f(t)$ does not become zero.
Then, the singularity in the BEC doesn't occurs in this case. This
is because of the existence of the harmonic trapping potential.
The harmonic potential which has zero point energy in the ground
state plays a central role in
preventing the condensate from being singularity.\\
In conclusion, although the atoms are loaded with the same gain
coefficients, the time dependent nature of the scattering length
changes the number of atoms transferred to the BEC from the
reservoir as can be seen from (8). The transformations introduced
here provides for an analytic way to investigate the effect of
Feshbach resonance on the number of atoms which are continuously
loaded or depleted by
loss.\\

\end{document}